\newcommand{\CLASSINPUTtoptextmargin}{0.7in}
\newcommand{\CLASSINPUTbottomtextmargin}{1in}
\documentclass[conference]{IEEEtran} 
\usepackage{setspace,amssymb,graphicx,enumerate,amsfonts,amsmath,color,algorithm,algpseudocode}
\usepackage{epsfig,subfigure,booktabs,amsmath,calc,arydshln,tikz,pdflscape,rotating}
\usepackage{cite}
\usepackage[english]{babel}
\usepackage{epstopdf,multirow}
\usetikzlibrary{shapes}
\usepackage{calc}
\usepackage{ifthen}
\usepackage{verbatim}

\IEEEoverridecommandlockouts

\begin{document}
\bstctlcite{IEEEexample:BSTcontrol}

\title{CLOEE - Cross-Layer Optimization for Energy Efficiency  of IEEE~802.15.6 IR-UWB WBANs}
\author{\IEEEauthorblockN{Kemal Davaslioglu, Yang Liu, and Richard D. Gitlin}
\IEEEauthorblockA{Department of Electrical Engineering\\
University of South Florida\\
Email: kemald@usf.edu, yangl@mail.usf.edu, richgitlin@usf.edu}}
\maketitle
\begin{abstract}
Advances in sensor networks enable pervasive health monitoring and patient-specific treatments that take into account the patient’s medical history, current state, genetic background, and personal habits. However, sensors are often battery power limited and vastly differ in their application related requirements. In this paper, we address both of these problems. Specifically, we study IEEE 802.15.6 ultra-wideband (UWB) wireless body area networks (WBANs) that use impulse radio (IR). We formulate the joint optimization of the payload size and number of pulses per symbol to optimize the energy efficiency under the minimum rate constraint. The proposed algorithm, cross-layer optimization for energy efficiency (CLOEE), enables us to carry out a cross-layer resource allocation that addresses the rate and reliability trade-off in the physical (PHY) layer as well as the packet size optimization and transmission efficiency for the medium access control (MAC) layer. Simulation results demonstrate that CLOEE can provide an order of magnitude improvement in energy efficiency or extend the range by a factor of two compared to static strategies.
\end{abstract}
\section{Introduction}
A plethora of sensor applications and wearable technologies exists that enable the ubiquitous recording and storage of vital health signs. With the advances in, and miniaturization of sensors such as implantable medical sensors  (i.e., \emph{in vivo} sensors), health and fitness trackers, health monitors, and smart clothing, patients can be monitored without hindering their daily activities. In order to address the unique demands of wireless body area networks (WBANs), the IEEE~802.15.6 standard finalized in 2012 defines a standard for a short-range, extremely low power (also low-duty cycle), and reliable communication to support a variety of applications for medical monitoring and personal entertainment \cite{IEEE802156}. The IEEE 802.15.6 standardizes three physical (PHY) layer methods: narrowband PHY, ultra wideband (UWB) PHY, and human body communications PHY. In this paper, we are interested in the applications using the UWB PHY instead of the narrowband PHY as UWB PHY is more robust to the channel variations and operates at very low-power levels for human body applications \cite{Capriotti02}. 

The application areas, channel models, standards, recent research efforts, and design challenges of WBANs have been studied in numerous surveys,  see, e.g., \cite{Touati13,Smith14,Cavallari14}. Our interest here is on the resource allocation for energy efficiency and throughput. In general, this area has been primarily investigated for the narrowband PHY applications, see e.g., \cite{Franco10,Tachtatzis12,Babu14,Chuang13}. This is probably due to an attractive feature that the standard introduces: a $m$-periodic scheduled allocation mode where the hub and nodes communicate in every $m$~superframes allowing the nodes to sleep between superframes. The optimal~$m$ that maximizes the device lifetime has been investigated in  \cite{Franco10}, where the current drawn in different states of the nodes are taken into account. The numerical study in \cite{Franco10} was later extended in  \cite{Tachtatzis12} by solving the problem by integer programming techniques. However, in both studies, the packets are assumed to be error-free and no closed-form expressions were obtained for the optimal MAC parameters. Note that, depending on the size of the problem, a mixed integer programming solution may be computationally expensive for a low-power device. The effects of erroneous transmissions and forward error correcting (FEC) for $m$-periodic scheduled allocation mode were numerically evaluated in \cite{Babu14}, but again without any closed-form expressions. Reference \cite{Chuang13} extended the results of \cite{Franco10,Tachtatzis12,Babu14} to two-hop relay nodes. 
 
Energy efficiency of impulse-radio (IR) UWB PHY for IEEE~802.15.6 applications has been recently studied in the literature, see, e.g., \cite{Karvonen14,Mohammadi14,Mohammadi14b}. These papers have slightly different definitions for energy efficiency. In  \cite{Karvonen14}, it is defined as the number of bits that can be successfully received per energy consumption with the units of $\mathrm{bits/Joule}$, whereas the one in \cite{Mohammadi14} divides the energy consumed to generate $l$ payload bits to the one for the payload plus overhead (unitless). Both metrics can be extended to the case that accounts for channel errors. A numerical evaluation is presented in \cite{Karvonen14} to find the optimal coding rate for different distances and modulations. The optimal frame length without any QoS constraints is determined in \cite{Mohammadi14} using a closed-form expression. The bit error probability of an IR-UWB system that accounts for the effects of intra-symbol interference is derived in \cite{Mohammadi14b}.

In this paper, we formulate an energy efficiency maximization problem for the IEEE 802.15.6 IR-UWB PHY. We propose a cross-layer optimization algorithm for energy efficiency (CLOEE) to determine the PHY and MAC layer parameters. Specifically, we focus on the optimal payload size and number of pulses per burst. The effects of FEC on the successful packet detection, QoS constraints of minimum throughput, and static power consumption in the circuitry are included in the formulation. We first prove that the energy efficiency is a quasiconcave function of the frame length. Then, we derive closed-form expressions for the optimal frame length that maximizes the energy efficiency and throughput, and propose a low-complexity algorithm to determine the optimal frame length and number of pulses per burst with constraints. We note that prior work in this area has considered either optimization with respect to only one of the parameters and without addressing the rate constraints \cite{Karvonen14,Mohammadi14,Mohammadi14b}, or when they did, they relied on high complexity integer programming solutions for narrowband PHY and no closed-form expressions were obtained \cite{Franco10,Tachtatzis12}. In this paper, we provide a comprehensive solution to address the shortcomings of the prior work and provide insight on determining where the crossovers of these parameters occur in order to facilitate real-time link adaptation.



The rest of this paper is organized as follows. Section~\ref{Section:SystemModel} introduces the system model. Section~\ref{Section:EE} describes the energy consumption model, energy efficiency maximization formulation, and proposed algorithm. In Section~\ref{Section:Simulations},  simulation results are presented and Section~\ref{Section:Conclusion} summarizes our conclusions. 
\section{System Model}\label{Section:SystemModel}
\subsection{Overview of the IEEE~802.15.6 UWB PHY}\label{Subsection:802156Std}
We consider a set of WBAN sensor nodes and a master hub node in which the medium access and power management functionalities are coordinated by the hub. According to the IEEE~802.15.6 standard, a hub can serve up to 64 nodes \cite{IEEE802156}. 
Eleven channels are defined for the UWB PHY in the 3.1-10.6~GHz spectrum band, each with a channel bandwidth of 499.2~MHz. 
There are three supported modulation schemes for IR-UWB, namely on-off modulation, differential binary phase shift keying (DBPSK) and differential quadrature phase shift keying (DQPSK). There are two modes of operation defined in \cite{IEEE802156} such as the default and the high QoS modes. The default mode is for the medical and non-medical applications, whereas the high QoS mode is to be used for high-priority medical applications only. In the sequel, we only discuss the default mode due to the space limitation.

\subsection{UWB PHY Superframe Structure}\label{Subsection:PDDU}
The UWB PHY frame format is referred to as the physical layer protocol data unit (PPDU). It is composed of the synchronization header (SHR), the physical layer header (PHR), and the physical layer service data unit (PSDU) as illustrated in Fig. 1. The PPDU duration is given by
\begin{align} 
T_{\text{packet}} = T_{\text{SHR}} + T_{\text{PHR}} + T_{\text{PSDU}},
\end{align}
where the SHR, PHR, and PSDU frame durations are denoted by $ T_{\text{SHR}}$, $ T_{\text{SHR}}$, and $T_{\text{PSDU}}$, respectively.

The SHR frame consists of two parts. The first part is the preamble and it is used for timing synchronization, packet detection, and carrier frequency offset recovery.
The preamble also enables the coexistence of WBANs \cite{IEEE802156}. The second part is the start-of-frame delimiter (SFD) for frame synchronization. The preamble and SFD are made up of four and one Kasami sequences of $63$ bits, respectively. Between the bits of a Kasami sequence, $L-1$ zeros are padded. To keep the duty cycle low, $L \cdot T_w$ is fixed to $128$~nsec \cite{IEEE802156}. We take $T_w = 8$~nsec, $L = 16$, and  $T_{\text{SHR}} = 5 \cdot 63 \cdot 128~\mathrm{ nsec } = 40.32$~$\mu$sec as in \cite{IEEE802156}. \label{Time:SHR}  

The PHR frame consists of 24~bits that carry information about the data rate of the PSDU, MAC frame body length, pulse shape, burst mode, HARQ, and scrambler seed. A shortened Bose-Chaudhuri-Hocquenghem (BCH) code of $(40,28;2)$ is used such that $k = 28$  bits are appended with parity bits to form codewords of length $n = 40$ bits with an error correcting capability of $t_{\text{ECC}} = 2$. Hence, the PHR frame bits, $N_{\text{PHR}} = 40$, are transmitted at a sampling rate of $2051.2$~nsec, i.e., $T_{\text{PHR}}  = 40 \times 2051.3$ nsec $= 82.052$~$\mu$sec.

\begin{figure}[t!]
\centering
\includegraphics[width=0.8\columnwidth]{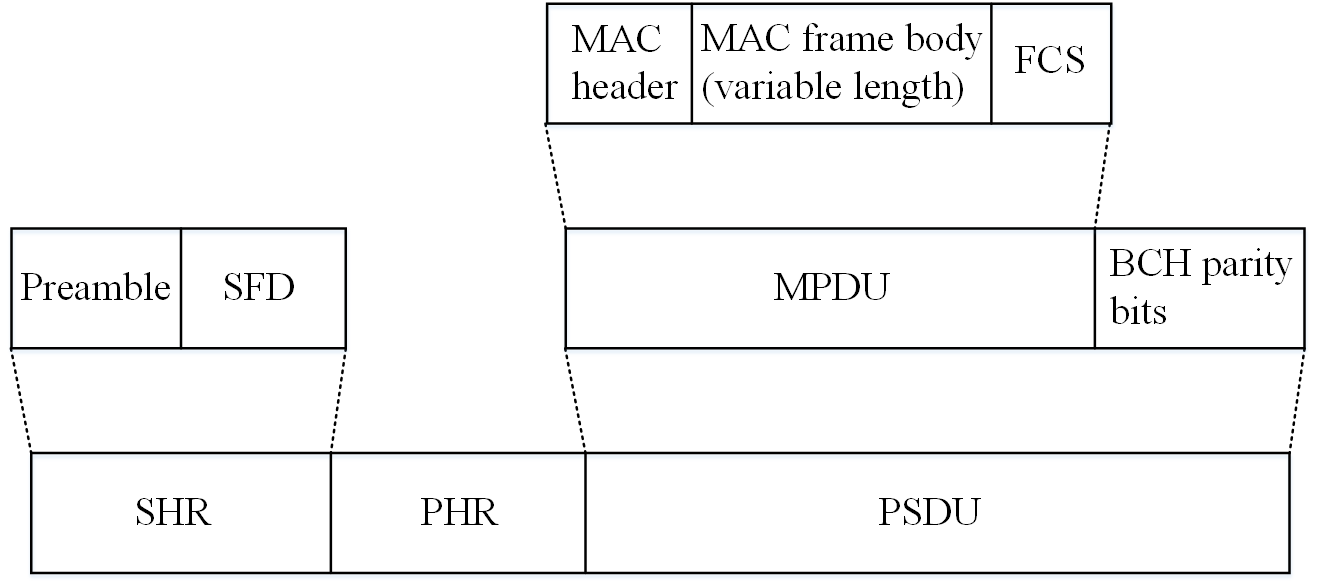}
\label{Figure:UWBPPDU}
\caption{IEEE 802.15.6 UWB PPDU frame structure.}  
\end{figure}

The PSDU includes the MAC protocol data unit (MPDU) and the channel code, BCH parity bits, in the default mode. The MPDU consists of a MAC header, a variable length MAC frame body, and a frame check sequence (FCS). The header and FCS of the MPDU frame are $N_{\text{FCS}} = 56$ and $N_{\text{MH}} = 16$ bits, respectively. The MAC frame body has a variable length of $N_{\text{FB}}^{'}$ bits. Thus, in a MPDU frame, the total number of bits before bit stuffing are given as $ N_{\text{MPDU}}^{'} = N_{\text{MH}} + N_{\text{FCS}} + N_{\text{FB}}^{'} $. These are grouped in blocks of  length $k$ to codewords of length $n$. In Table~\ref{Table:ECC}, the error correcting codes (ECC) for the PHR and PSDU frames are summarized for the default and high QoS modes, where we follow the notation $(n,k;t_{\text{ECC}})$ for the BCH codes. To align the symbol boundaries, bits are padded to the last word if $\mathrm{rem}(N_{\text{FB}}^{'} + 72 , k ) \neq 0$, where $\mathrm{rem}(x,y)$ is the remainder of $x$ divided by $y$. In that case, the last codeword would require $N_{\text{bs}} = N_{\text{CW}} k - N_{\text{MPDU}}^{'}$ bit stuffing such that the total number of bits before encoding becomes $N_{\text{MPDU}} = N_{\text{MPDU}}^{'} + N_{\text{bs}}$. When the BCH parity bits are included, the total number of payload bits for the PSDU frame becomes
\begin{align}
N_{\text{T}} = N_{\text{MPDU}} + (n-k) N_{\text{CW}}.
\end{align} 
The duration of the PSDU frame is $T_{\text{PSDU}} = N_{\text{T}}/ R_b$, where $R_b$ is one of the uncoded bit rates in Table~\ref{Table:DataRate}. The problem formulation in Section~\ref{Subsection:ProblemFormulation} will use the $N_{\text{T}}$ to relate the error probability which depends on the number of code words $N_{\text{CW}}$. Since $N_{\text{MH}} + N_{\text{FCS}} = 72$~bits, we can express $N_{\text{CW}}$ as
\begin{align} N_{\text{CW}} = \left\lceil \frac{N_{\text{MPDU}} + 72}{k} \right\rceil \approx \left\lceil \frac{N_{\text{T}}}{n} \right\rceil,
\end{align}
where the smallest integer greater than or equal to a real number $x$ is given by $\lceil x \rceil$. In Section~\ref{Section:Simulations}, we observe that this approximation results in a negligible loss. Relating $N_{\text{CW}}$ to $N_{\text{T}}$ will help us derive the expressions in Section~\ref{Subsection:ECC}.



\begin{table}[t!]
\caption{BCH$(n,k;t_{\text{ECC}})$ error correcting codes of the PHR and PSDU frames in the IEEE 802.15.6 UWB PHY}
\label{Table:ECC}
\centering
\footnotesize{\begin{tabular}{ccc}\toprule
Frame  Type & Default mode & High QoS Mode \\\midrule
PHR & BCH(40,28;2)  & BCH(91,28;10) \\
PSDU & BCH(63,51;2) & BCH(126,63;7) \\
 \bottomrule
\end{tabular}}
\end{table}

\subsection{Modulation, Receiver, and Probability of Bit Error}\label{Section:Modulation}

On-off modulation is a combination of M-ary waveform coding and on-off keying. With the on-off modulation, $K$ bits from an alphabet size of $M = 2^K$ are grouped, $(b_0,b_1,\cdots,b_{K-1})$, and passed through a symbol mapper of rate $1/2$ such that an output sequence of $2K$ bits, $(d_0,d_1,\cdots,d_{2K-1})$, is obtained that has the same alphabet size. 
In \cite{IEEE802156}, $K = 1$ is considered as the default mode with an optional mode of $K = 4$. For example, the input bit $0$ is mapped to $[1 \hspace{0.2em} 0]$, whereas $1$ is mapped to $[0 \hspace{0.2em} 1]$. Note that the performance of the on-off modulation for $K = 1$ closely follows the one for binary pulse position modulation \cite[Appendix~C]{IEEE802156}. After processing by the pulse shaping filter, the $m$th output IR-UWB symbol can be expressed as \cite{IEEE802156}
\begin{align}\label{Eqn:xm}
x_{m}(t) = \sum_{n = 0}^{2K-1} d_n^{(m)} w_{2Km+n} & \left(t -  n(T_{\text{sym}}/2) - mKT_{\text{sym}}  \right. \nonumber \\
& \hspace{1em} \left. - h_{2Km+n}T_w \right), 
\end{align}
where $d_n^{(m)}$ is the $n$th codeword component of the $m$th symbol, $T_{\text{sym}}$ is the symbol time, and $\{ h_{2Km+n} \}$ is the time-hopping sequence. The symbol time $T_{\text{sym}}$ has $N_w$ pulse waveform positions each with a duration of $T_w$,  $T_{\text{sym}} = N_w T_w$. The symbol duration is divided into two intervals of duration $T_{\text{sym}}/2$ in order to enable on-off modulation. The duty cycle $T_w / T_{\text{sym}}$ is fixed at $1/32 = 3.125\%$ in \cite{IEEE802156} to ensure low power consumption. The pulse waveform $w_n(t)$ is given by
\begin{align}w_n(t) = \sum_{i = 0}^{N_{\text{cpb}}-1} ( 1 - 2 s_{i} ) p\left(t - i T_p \right), \end{align} where $p(t)$ denotes a single pulse of duration $T_p$ \cite{IEEE802156}.
The sequence $\{ s_i\}$ denotes the scrambling sequence that helps reduce the spectral lines due to same polarity pulses \cite{IEEE802156,Gezici05}. The integer $N_{\text{cpb}}$ defines the number of pulses per burst and $N_{\text{cpb}} \geq 1$. In the single pulse case, $N_{\text{cpb}} = 1$, whereas $N_{\text{cpb}} \in \{2,4,8,16,32 \}$ for the burst pulse option. Note that the processing gain of an IR-UWB system is  $N_{\text{cpb}}N_w$ \cite{Gezici05}. Since $N_{\text{w}}$ is fixed by the standard \cite{IEEE802156}, the processing gain can be varied by changing $N_{\text{cpb}}$. 

Table~\ref{Table:DataRate} presents the timing parameters and data rates for different transmission modes for on-off modulation. The pulses are generated at a frequency of $499.2$~MHz and $T_p$ is $2.0032$~nsec. The symbol is encoded with a BCH code of (63,51).  An uncoded symbol rate, $R_b  = 1/T_{\text{sym}}$, is multiplied with the FEC rate to obtain the coded symbol rate. Hence, $N_{\text{cpb}}$ is used to balance the data rate and processing gain trade-off. 



A non-coherent detector is considered and equally likely input bits are assumed. Each pulse has an energy of $\varepsilon_p = \varepsilon_b/N_{\text{cpb}}$. The bit error probability is given by \cite{Witrisal09,Karvonen14} 
\begin{align} \label{Eqn:Pb}
P_b = Q\left(\sqrt{\frac{1}{2} \cdot \frac{(h\varepsilon_b/N_0)^2}{h \varepsilon_b / N_0 + N_{\text{cpb}} T_{\text{int}} W_{\text{rx}}}}\right),  
\end{align}
where $h$ is the channel coefficient, $T_{\text{int}}$ is the integration interval per pulse, $W_{\text{rx}}$ is the equivalent noise bandwidth of the receiver front end, and $\varepsilon_p/N_0$ is the integrated signal-to-noise ratio per bit. In Section~\ref{Section:Simulations}, we take the integration time as the pulse duration, i.e., $T_{\text{int}} = N_{\text{cpb}} T_p$, and assume that the receiver and transmitter are fully synchronized. 



\begin{table}[t!]
\caption{Data rate and symbol timing related parameters of the IEEE~802.15.6 UWB PHY \cite{IEEE802156}}
\label{Table:DataRate}
\centering  
\footnotesize{\begin{tabular}{ccccc}
\toprule
\multirow{2}{*}{$N_{\text{cpb}}$} &$T_w$ & $T_{\text{sym}}$ & Uncoded Bit  & Coded Bit \\
 & (nsec)  &  (nsec)  & Rate  (Mbps)& Rate (Mbps)  \\\midrule
 $32$ & $64.103$ & $2051.3$ & $0.488$ & $0.395$  \\
 $16$ & $32.051$ & $1025.6$ & $0.975$ & $0.790$ \\ 
 $8$ & $16.206$ & $512.8$ & $1.95$0 & $1.580$  \\ 
 $4$ & $8.012$ & $256.4$ & $3.900$ & $3.159$  \\
 $2$ & $4.006$ & $128.2$ & $7.800$ & $6.318$  \\
 $1$ & $2.003$ & $64.1$ & $15.600$ & $12.636$  \\\bottomrule
\end{tabular}}
\end{table}


\subsection{Error Correction}\label{Subsection:ECC} 
In Section~\ref{Subsection:PDDU}, we discussed the structure of an UWB frame that consists of SHR, PHR, and PSDU, and in Section~\ref{Section:Modulation}, the bit error probability is presented for a non-coherent ED receiver. In what follows, we discuss the error correcting capabilities of each frame type. 

The SHR frame is correctly received at the receiver if both the preamble and the SFD transmissions are successful, which can be mathematically expressed as 
\begin{align}\label{Eqn:PSHR} P_{\text{SHR}} = P_{\text{SFD}} (1 - (1 - P_{\text{Kasami}} )^4 ), \end{align}
where $P_{\text{SFD}}$ and $P_{\text{Kasami}}$ are the probabilities of correctly decoding the SFD and Kasami sequence, respectively. Since there are four Kasami sequences in the preamble, we have $\left(1 - P_{\text{Kasami}} \right)^4$ in (\ref{Eqn:PSHR}). The probability of successful delivery of a 63-bit Kasami sequence can be expressed as \cite{Mohammadi14} $P_{\text{Kasami}} = \sum_{i = 0}^{\rho} \binom{63}{i} (P_b)^i (1-P_b)^{63 - i}$, where the operator $\binom{a}{b}$ represents the binomial coefficient and $\rho$ is an implementation-dependent sensitivity margin and it is taken as $\rho = 6$ as in \cite{Mohammadi14}. Since the SFD is the ones complement of a Kasami sequence, we have $P_{\text{SFD}} = P_{\text{Kasami}}$. 

The BCH decoder can recover up to $t_{\text{ECC}}$ bit errors for a BCH(n,k;t) code. Then, for the PHR frame, the probability of successful reception of a codeword is \cite{Gitlin92}
\begin{align} P_{\text{PHR}}  = \sum\limits_{i = 0}^{t} \binom{N_{\text{PHR}}}{i} (P_b)^i (1-P_b)^{N_{\text{PHR}} - i}, \end{align}
where $N_{\text{PHR}} = 40$~bits, and $P_b$ and $t_{\text{ECC}}$ are given in (\ref{Eqn:Pb}) and Table~\ref{Table:ECC}, respectively. 

The PSDU frame consists of $N_{\text{CW}}$ codewords. The probability of successful reception of PSDU frame is if all the codewords are received successfully, that is
\begin{align} P_{\text{PSDU}} = (P_{\text{CW}})^{N_{\text{CW}}} = (P_{\text{CW}})^{\left\lceil \frac{N_{\text{T}}}{n} \right\rceil},  \end{align}
where $P_{\text{CW}}$ is the probability successful reception of a codeword that can be expressed as 
\begin{align} P_{\text{CW}} = \sum\limits_{i = 0}^{t} \binom{n}{i} (P_b)^i  (1-P_b)^{n - i}. \end{align}

The probability of  successful delivery is when the SHR, PHR, and PSDU frames are successfully received, that is 
\begin{align} P_{\text{PPDU}}  = P_{\text{SHR}}  P_{\text{PHR}}  P_{\text{PSDU}} = P_{\text{SHR}}  P_{\text{PHR}} ( P_{\text{CW}} )^{\left\lceil \frac{N_{\text{T}}}{n} \right\rceil} \end{align}
where $P_{\text{SHR}}$, $P_{\text{PHR}}$,  and $P_{\text{PSDU}}$ are the successful delivery probabilities of the each frame type, respectively. For the default mode, $k = 51$ as shown in Table~\ref{Table:DataRate}. On one hand, frame error probability increases as $N_{\text{T}}$ increases, while, on the other hand, a short $N_{\text{T}}$ will result in system inefficiency due to high packet overhead. 

\section{Energy Efficiency Maximization}\label{Section:EE} 
\subsection{Energy Consumption Model}\label{Subsection:EnergyCModels}
There have been several models in the literature  characterizing the energy consumption of IR-UWB radios, e.g., \cite{Seyedi10,Mercier15,Mohammadi14,Karvonen14}.  Among these studies, we use the model in \cite{Seyedi10} as it provides a general model to accommodate  the coherent and non-coherent detectors, hard and soft decision demodulators, and different modulation types. 
Using this model, the energy required to transmit and receive a payload bit is given by
\begin{align}\label{Eqn:E-b}
\varepsilon_B = (\varepsilon_{\text{FB}}^{\text{Tx}} + \varepsilon_{\text{FB}}^{\text{Rx}})/N_{T},
\end{align}
where $\varepsilon_{\text{FB}}^{\text{Tx}}$ and $\varepsilon_{\text{FB}}^{\text{Rx}}$ represent the energy consumed at the transmitter and receiver for the PSDU frame, respectively. These two terms can be expressed as 
\begin{align}\label{Eqn:E-fb}
&\varepsilon_{\text{FB}}^{\text{Tx}} = \varepsilon_p N_{\text{cpb}} N_{T}   + P_{\text{SYN}} T_{\text{onL}}, \\
&\resizebox{0.95\columnwidth}{!}{$\varepsilon_{\text{FB}}^{\text{Rx}} =  ( MP_{\text{COR}} + \rho_c P_{\text{ADC}} + P_{\text{LNA}} + P_{\text{VGA}} + \rho_r (P_{\text{GEN}} + P_{\text{SYN}})) T_{\text{onL}}$}, \nonumber
\end{align}
where $P_{\text{SYN}}$ is the power consumption of the clock generator and synchronizer at the transmitter and $T_{\text{onL}}$ is the time duration to transmit $N_T$ bits, $T_{\text{onL}} = T_{\text{sym}} N_T$. The terms $P_{\text{COR}}$, $P_{\text{ADC}}$, $P_{\text{LNA}}$, $P_{\text{VGA}}$, and $P_{\text{GEN}}$, respectively, represent the power consumption of the RAKE fingers of the receiver, the analog-to-digital converter (ADC), the low noise amplifier (LNA), the variable gain amplifier (VGA), and the pulse generator. The number of RAKE receiver fingers is denoted by $M$, the term $\rho_r = 1$ for coherent modulation, and $\rho_r = 0$ for non-coherent modulation, and  the term $\rho_c = 1$ for soft decision, whereas $\rho_c = 0$ is for hard decision.

Similarly, the overhead energy consumption is defined as $\varepsilon_{\text{OH}}  = \varepsilon_{\text{OH}}^{\text{Tx}} +  \varepsilon_{\text{OH}}^{\text{Rx}}$, where $\varepsilon_{\text{OH}}^{\text{Tx}}$ and $\varepsilon_{\text{OH}}^{\text{Rx}}$ denote the energy to transmit and receive the overhead, respectively, and these are given by 
\begin{align}\label{Eqn:oh} \varepsilon_{\text{OH}}^{\text{Tx}} =&(N_{\text{cpb}}^{\text{SHR}} N_{\text{SHR}} + N_{\text{cpb}}^{\text{PHR}} N_{\text{PHR}}) \varepsilon_p + P_{\text{SYN}} (T_{\text{SHR}} + T_{\text{PHR}}), \end{align}
\begin{align}\begin{aligned}
\varepsilon_{\text{OH}}^{\text{Rx}} =&   \left( MP_{\text{COR}} + \rho_c P_{\text{ADC}} + P_{\text{LNA}} + P_{\text{VGA}} \right.\\
& + \left. \rho_r (P_{\text{GEN}} + P_{\text{SYN}}) \right) ( T_{\text{SHR}} +  T_{\text{PHR}} ),
\end{aligned}
\end{align}
where $N_{\text{cpb}}^{\text{SHR}} = 4$, $N_{\text{cpb}}^{\text{PHR}} = 32$, $N_{\text{SHR}} = 63 \cdot 5 =315$,  and $T_{\text{SHR}}$ and $T_{\text{PHR}}$ are defined as in Section~\ref{Time:SHR}. 

Finally, the startup energy is $\varepsilon_{\text{ST}} = \varepsilon_{\text{ST}}^{\text{Tx}} + \varepsilon_{\text{ST}}^{\text{Rx}} = 2 P_{\text{SYN}} T_{\text{ST}}$,
where $T_{\text{ST}}$ is the time duration for the start up of the devices. 

\newtheorem{theorem2}{Theorem}
\subsection{Problem Formulation}\label{Subsection:ProblemFormulation}
Consider a hub and $N_{\text{S}}$ nodes, each requesting $R_0$  bits per second, in a one-hop star topology. The following are defined to aid the problem formulation:
\newtheorem{lemma}{Lemma}
\newtheorem{definition}{Definition}
\begin{definition} A function $f$ is strictly quasiconcave if its domain $\mathcal{D}$ is convex, and for any $x, y \in \mathcal{D}$ with $f(x) \neq f(y)$, the following is true for all $\lambda \in (0,1)$ \cite{NonlinearBazaraa}
\begin{align} f(\lambda x + (1-\lambda) y) > \min\{f(x),f(y) \}. 
\end{align}
\end{definition}
\begin{definition}\label{Definition:EE}
The energy efficiency is defined as the ratio of the total number of successfully received bits to the total energy consumed at the transmitter and receiver. It can be expressed in the units of $\mathrm{bits/Joule}$ as
\begin{align}\label{Eqn:Eta}
\eta(N_{\text{T}}, N_{\text{cpb}}) = \frac{N_\text{T}  P_{\text{SHR}}  P_{\text{PHR}} (P_{\text{CW}})^{\left\lceil \frac{N_{\text{T}}}{n}  \right\rceil}}{  N_{\text{T}} \cdot \varepsilon_{\text{B}} + \varepsilon_{\text{OH}} + \varepsilon_{\text{ST}} }.
\end{align} 
\end{definition}
\begin{lemma} The optimal PSDU frame size that maximizes (\ref{Eqn:Eta}) can be expressed as
\begin{align}\label{Eqn:EEwoConstraints}
\small{N_{\text{T}}^{\text{EE}} = \left[ \sqrt{ \frac{(\varepsilon_{\text{OH}} + \varepsilon_{\text{ST}})^2}{(2 \varepsilon_{\text{B}})^2}  -  \frac{n ( \varepsilon_{\text{OH}}+\varepsilon_{\text{ST}})}{\varepsilon_{\text{B}} \log(P_{\text{CW}})}} - \frac{\varepsilon_{\text{OH}} + \varepsilon_{\text{ST}} }{2 \varepsilon_{\text{B}}} \right]}
\end{align}
\end{lemma}
\begin{IEEEproof} It follows that when we take the derivative of (\ref{Eqn:Eta}) with respect to $N_{\text{T}}$ and rearrange the terms, it is straightforward to obtain  (\ref{Eqn:EEwoConstraints}).
\end{IEEEproof}
\begin{theorem2} \label{Eqn:TheoremEEQuasi} Energy efficiency is strictly quasiconcave in $N_{\text{T}}$.  \end{theorem2}
\begin{IEEEproof} See the Appendix.  \end{IEEEproof}
\begin{theorem2}
If a function $\eta$ is a strictly quasiconcave, then a local optimal solution is also a global maximum solution.
\end{theorem2}
\begin{IEEEproof}
See the proof of Theorem~3.5.6 in \cite{NonlinearBazaraa}. 
\end{IEEEproof}

\begin{definition}\label{Definition:Throughput} The network throughput is defined as the ratio of the total number bits that are successfully received at receiver to the duration of a frame, and can be defined in the units of $\mathrm{bits/sec}$ as
\begin{align}\label{Eqn:Throughput}
R(N_{\text{T}}, N_{\text{cpb}})= \frac{N_\text{T}  P_{\text{SHR}}  P_{\text{PHR}} (P_{\text{CW}} )^{\left \lceil \frac{N_{\text{T}}}{n} \right\rceil} }{  T_{\text{SHR}} + T_{\text{PHR}} + N_{\text{T}} T_{\text{sym}}},
\end{align}
where the terms $P_{\text{SHR}}$ and $P_{\text{PHR}}$ are functions of $N_{\text{cpb}}$ through $P_b$. The probability $P_{\text{CW}}$ is a function of both $N_{\text{T}}$ and $N_{\text{cpb}}$.
\end{definition}
\begin{lemma} For a given $N_{\text{cpb}}$, the optimal PSDU frame size that maximizes the throughput, $R(N_{\text{T}},N_{\text{cpb}})$, is given by
\end{lemma}
\begin{align}\label{Eqn:ThrMaxSolution}
\resizebox{\columnwidth}{!}{$N_{\text{T}}^{\text{THR}} = \left[ \sqrt{ \frac{(T_{\text{SHR}} + T_{\text{PHR}})^2}{(2 T_{\text{sym}})^2}  -  \frac{n ( T_{\text{SHR}}+T_{\text{PHR}})}{T_{\text{sym}} \log(P_{\text{CW}})}} - \frac{T_{\text{SHR}} + T_{\text{PHR}} }{2 T_{\text{sym}}} \right]$}.
\end{align} 
\begin{IEEEproof}
When the derivative of (\ref{Eqn:Throughput}) is taken with respect to $N_{\text{T}}$, equate it to zero, and rearrange the terms, we obtain the expression in (\ref{Eqn:ThrMaxSolution}) for the optimal PSDU frame size.
\end{IEEEproof}



Our objective is to maximize the network energy efficiency subject to the minimum rate constraint, which is given by 
\begin{align}\label{Eqn:EEMax} \begin{aligned} 
\textbf{(P)}~\max & \hspace{0.75em} \frac{f(N_{\text{T}},N_{\text{cpb}})}{g(N_{\text{T}},N_{\text{cpb}})} =  \frac{N_\text{T} P_{\text{SHR}}  P_{\text{PHR}} ( P_{\text{CW}} )^{ \frac{N_{\text{T}}}{n}  } }{ (N_{\text{T}} \varepsilon_{\text{B}}+ \varepsilon_{\text{OH}} + \varepsilon_{\text{ST}})} \\
\mathrm{s. t.} &  \hspace{0.75em} \frac{N_\text{T} P_{\text{SHR}}  P_{\text{PHR}} ( P_{\text{CW}} )^{ \frac{N_{\text{T}}}{n} } }{ T_{\text{SHR}} + T_{\text{PHR}} + T_{\text{sym}} N_{\text{T}} }  \geq R_0 N_{\text{S}}. 
\end{aligned}
\end{align} 
The Lagrangian of (\ref{Eqn:EEMax}) can be expressed as
\begin{align} \label{Eqn:Lagrangian} 
\mathcal{L} (N_{\text{T}}, N_{\text{cpb}},\lambda) =& \eta(N_{\text{T}}, N_{\text{cpb}}) + \lambda (R(N_{\text{T}}, N_{\text{cpb}}) - R_0 N_{\text{S}} ), \end{align} 
where $\lambda$ is the Lagrangian variable associated with the minimum rate constraints.

\begin{algorithm}[t!]
  \caption{CLOEE -- Cross-Layer Optimization for Energy Efficiency for IEEE~802.15.6 IR-UWB}\label{Algorithm}
      {\begin{algorithmic}[1]
	\Function{Cloee}{$\varepsilon_{\text{B}},\varepsilon_{\text{OH}},\varepsilon_{\text{ST}},T_{\text{SHR}},T_{\text{PHR}},T_{\text{p}},R_0, N_{\text{S}}$}
 	\State Set an all-zeros vector $\textbf{N}_{\text{T}}$ of size $|\{N_{\text{cpb}}\}|$ and $n \gets 0$
	\For{$\textbf{N}_{\text{cpb}} = \{1,2,4,8,16,32\}$}
	\State Set $l \gets 0$ and solve (\ref{Eqn:EEwoConstraints}) to obtain $N_{\text{T}}(l)$
	\State Set $\lambda(l) \gets \max(R_0 N_{\text{S}} - R(N_{\text{T}}(l),\textbf{N}_{\text{cpb}}(n)) ,0)$ 
	\If{$R(N_{\text{T}}(l),\textbf{N}_{\text{cpb}}(n)) \geq R_0 N_{\text{S}}$}
	\State $\textbf{N}_{\text{T}}^{}(n) \gets N_{\text{T}}(l)$ 
	\Else
	\If{$R(N_{\text{T}}^{\text{THR}},\textbf{N}_{\text{cpb}}(n)) > R_0 N_{\text{S}}$}
	\Repeat	
	\State Solve (\ref{Eqn:DualProblem}) to obtain $N_{\text{T}}(l)$
	\State Update $\lambda(l+1)$ using (\ref{Eqn:DualUpdates}) 
	\State Set $l \gets l + 1$
	\Until{stopping criteria is satisfied} \label{Step:Termination}
	\State $\textbf{N}_{\text{T}}^{}(n) \gets N_{\text{T}}(l)$ 
	\Else
	\State $\textbf{N}_{\text{T}}^{}(n) \gets N_{\text{T}}^{\text{THR}}$ 
	\EndIf	
	\EndIf
	\State $n \gets n + 1$
	\EndFor
	\State  $(N_{\text{T}}^{\ast},N_{\text{cpb}}^{\ast}) \gets \arg\max_{(N_{\text{T}},N_{\text{cpb}})} \eta(\textbf{N}_{\text{T}}^{},\textbf{N}_{\text{cpb}})$ 
	\EndFunction
  \end{algorithmic}}
\end{algorithm}

\begin{lemma} The optimal solution, $(N_{\text{T}}^{\ast},N_{\text{cpb}}^{\ast})$, and the corresponding Lagrangian dual variable $\lambda^\ast$ must satisfy the following 
Karush-Kuhn-Tucker (KKT) conditions \cite{NonlinearBazaraa}
\begin{subequations}
\begin{equation}\label{Eqn:PrimalE1} \nabla_{N_{\text{T}}} \eta(N_{\text{T}}^{\ast},N_{\text{cpb}}^{\ast})  + \lambda^{\ast} \nabla_{N_{\text{T}}} R(N_{\text{T}}^{\ast},N_{\text{cpb}}^{\ast})  = 0,  \end{equation}
\begin{equation}\label{Eqn:PrimalE2} \nabla_{N_{\text{cpb}}} \eta(N_{\text{T}}^{\ast},N_{\text{cpb}}^{\ast})  + \lambda^{\ast} \nabla_{N_{\text{cpb}}} R(N_{\text{T}}^{\ast},N_{\text{cpb}}^{\ast}) = 0,  \end{equation}
\begin{equation}\label{Eqn:Slack1} \lambda^\ast [R(N_{\text{T}}^\ast, N_{\text{cpb}}^\ast) - R_0 N_{\text{S}} ] = 0, \text{ and } \lambda^{\ast} \geq 0, \end{equation}
\end{subequations} 
\end{lemma}
where $\nabla_x$ denotes the gradient with respect to $x$. Any point that satisfies (\ref{Eqn:PrimalE1})-(\ref{Eqn:PrimalE2}) is called a stationary point. Complementary slackness and dual feasibility conditions are expressed in (\ref{Eqn:Slack1}). These conditions suggest that if the minimum rate constraint is satisfied, $R(N_{\text{T}} , N_{\text{cpb}}) > R_0 N_{\text{S}}$, then $\lambda^{\ast} = 0$. Otherwise, we have $\lambda^{\ast} > 0$. 

\begin{figure}[t!]
\centering
\begin{tabular}{c}
		\subfigure[]{\includegraphics[width=0.9\columnwidth]{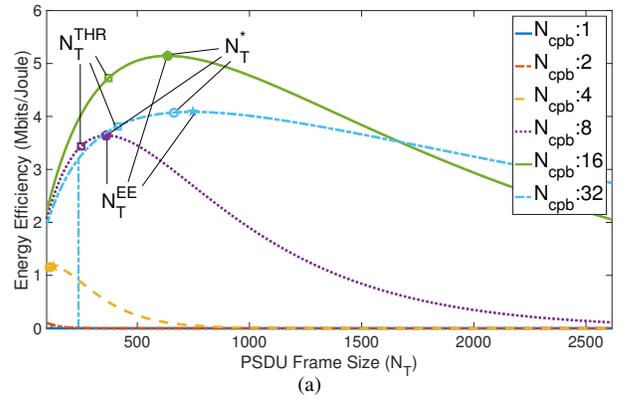}}\\
		\subfigure[]{\includegraphics[width=0.9\columnwidth]{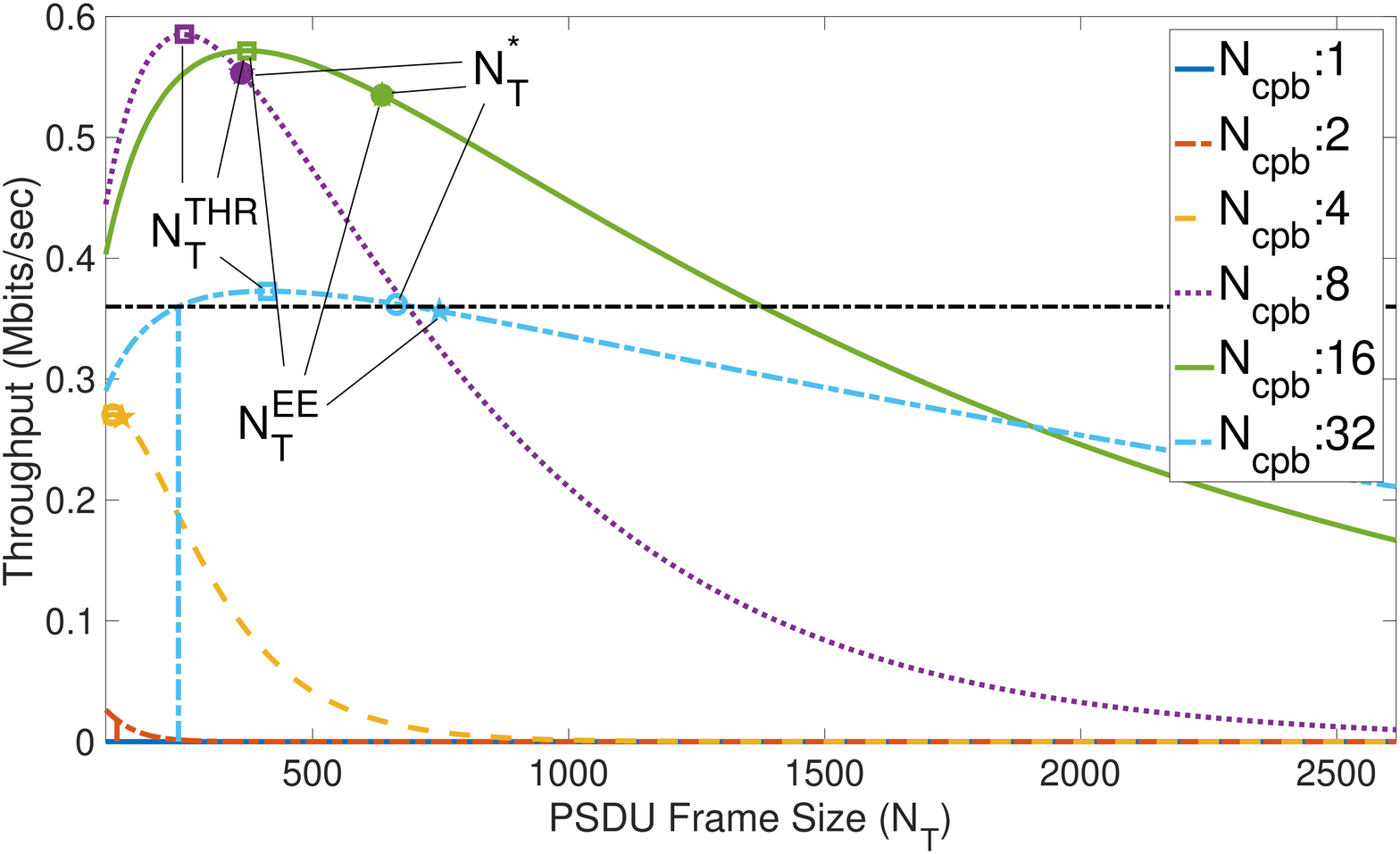}}\\
\end{tabular}
\caption{Energy efficiency and throughput versus the PSDU frame size at $8.4$~meters for $R_0 = 15$~kbits/sec and $N_{\text{S}} = 24$~nodes.}
\label{Figure:EEThrVSPayload}
\end{figure}


\begin{figure*}[t!]
\centering
\begin{tabular}{cccc}			
		\subfigure[]{\includegraphics[width=0.325\textwidth]{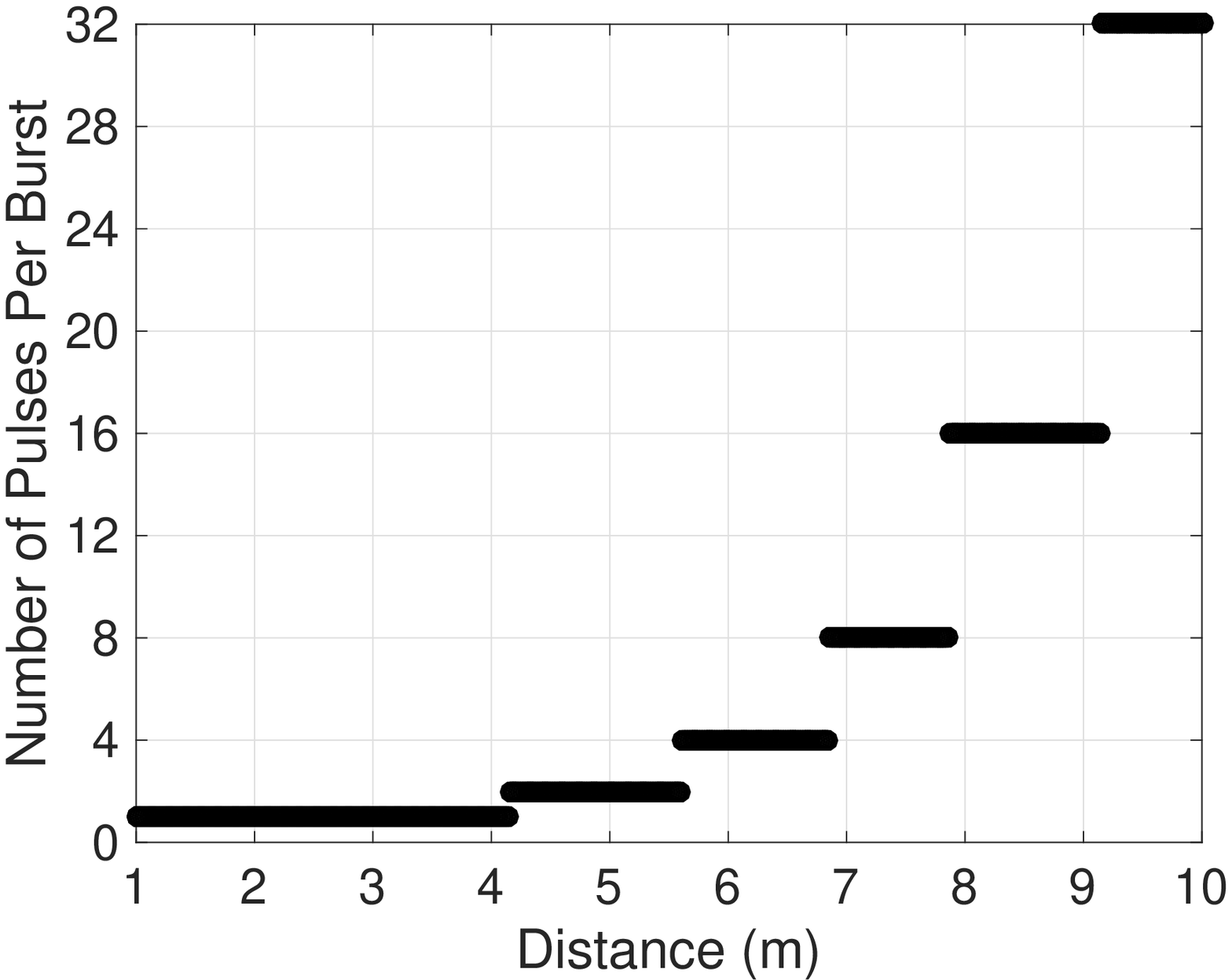}}&
		\subfigure[]{\includegraphics[width=0.325\textwidth]{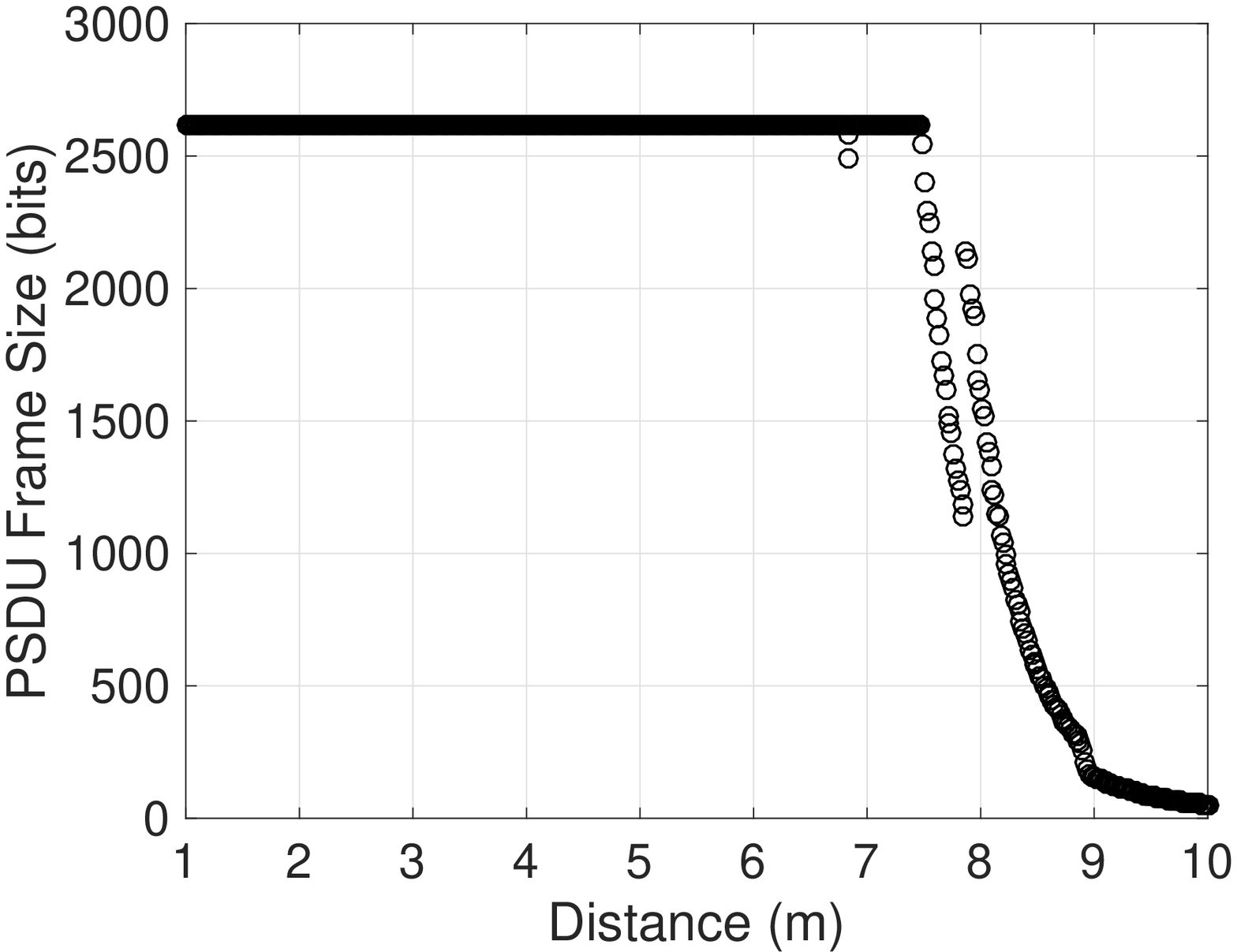}}\\
		\subfigure[]{\includegraphics[width=0.325\textwidth]{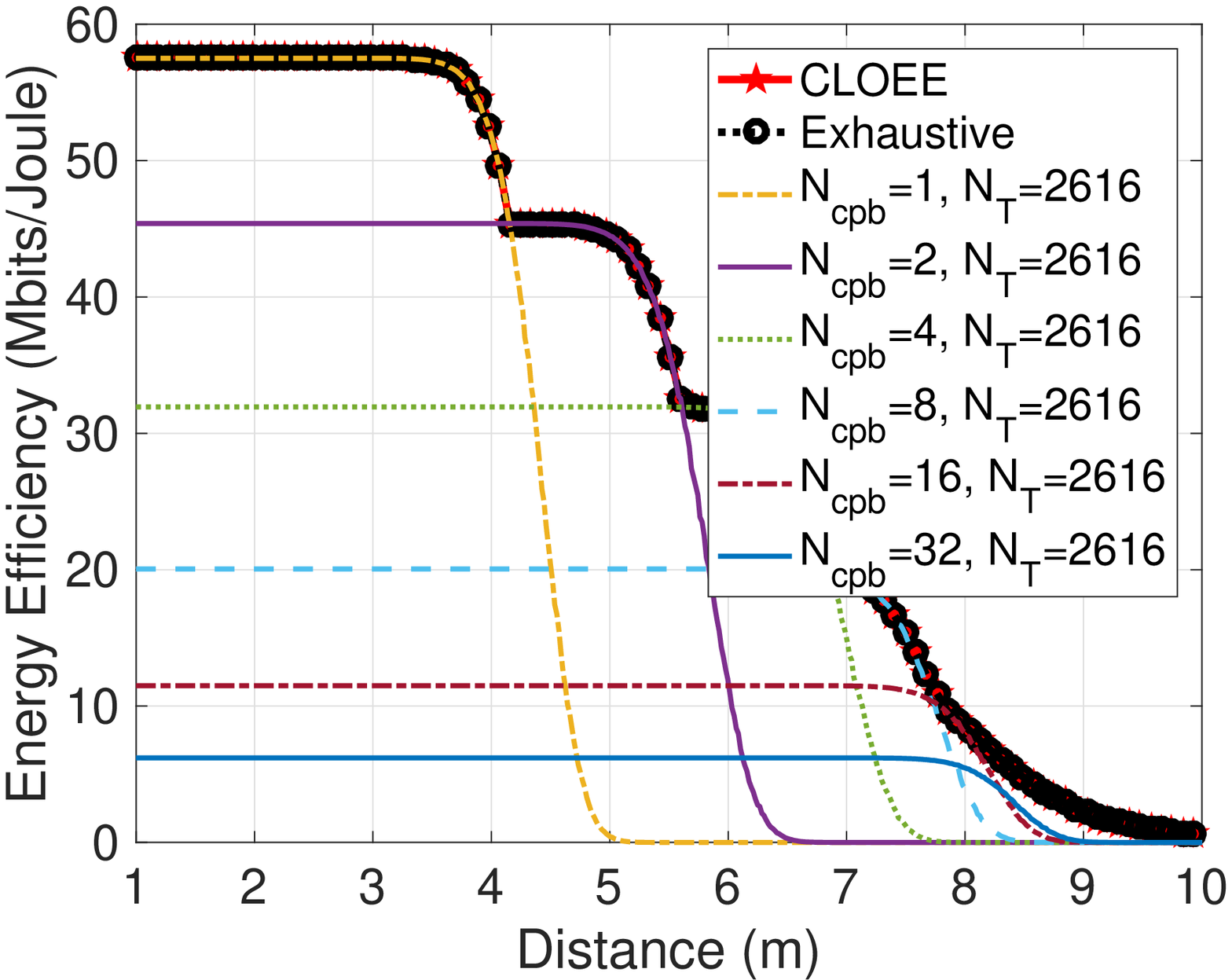}}&
		\subfigure[]{\includegraphics[width=0.325\textwidth]{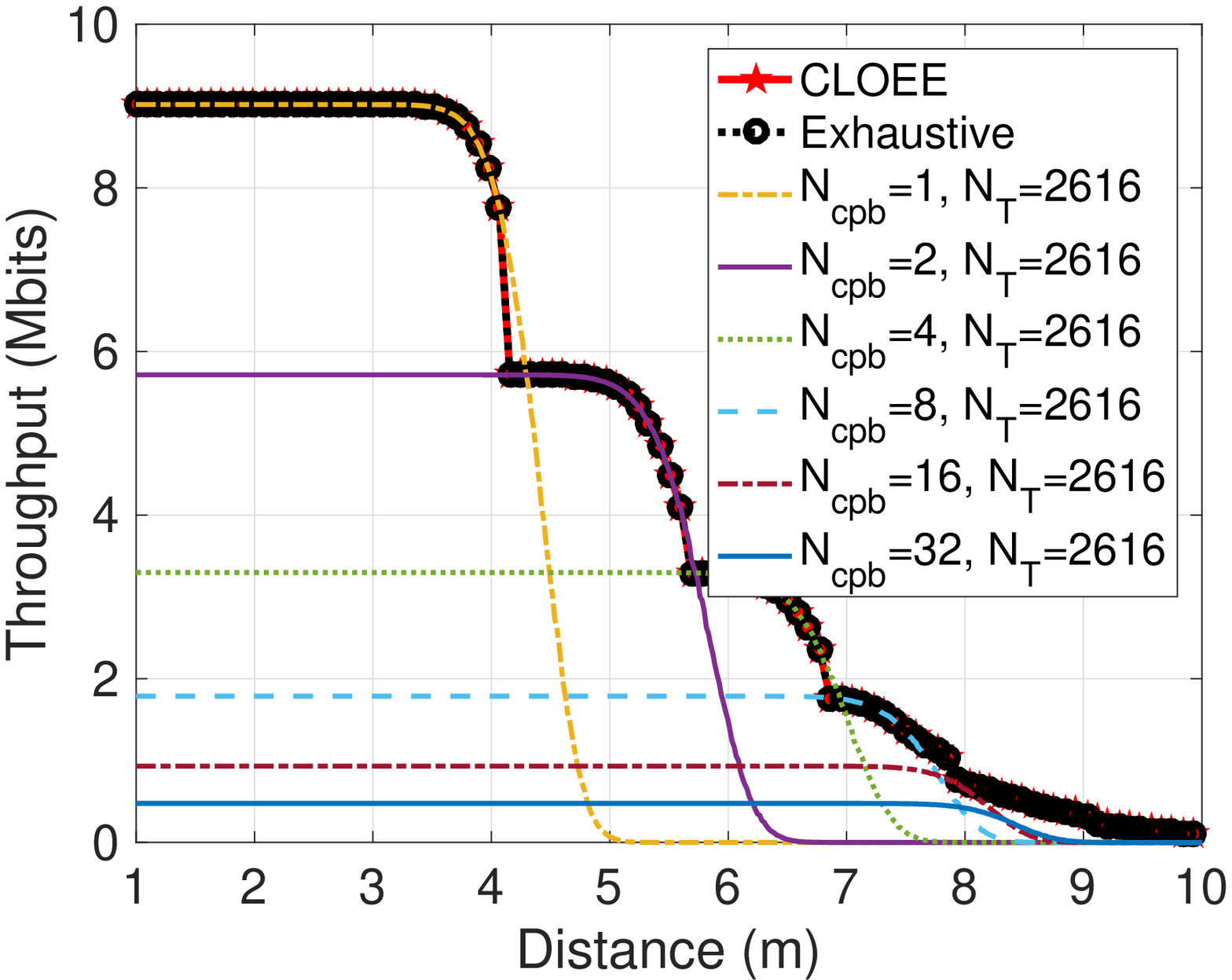}}				\end{tabular}
\caption{Link adaptation results for maximizing the energy efficiency.}  
\label{Figure:DistanceVsKPI}
\end{figure*}

The problem (\ref{Eqn:EEMax}) is a single-ratio fractional program and it can be solved using the dual fractional program \cite{HandbookGO}. Let $h(N_{\text{T}},N_{\text{cpb}})$ denote the constraint of (\ref{Eqn:EEMax}) such as
\begin{align}
h(N_{\text{T}},N_{\text{cpb}}) = R_0 N_{\text{S}} - \frac{N_\text{T} P_{\text{SHR}}  P_{\text{PHR}} ( P_{\text{CW}} )^{ \frac{N_{\text{T}}}{n} } }{ T_{\text{SHR}} + T_{\text{PHR}} + T_{\text{sym}} N_{\text{T}}}.
\end{align}
Then, the dual fractional program can be expressed as \cite{HandbookGO}
\begin{align} \label{Eqn:DualProblem}
\textbf{(D)}~ \min_{\lambda \geq 0} \left[ \max_{N_{\text{T}}} \frac{f(N_{\text{T}},N_{\text{cpb}}) - \lambda h(N_{\text{T}},N_{\text{cpb}})}{g(N_{\text{T}},N_{\text{cpb}})} \right].
\end{align}
Problem (\ref{Eqn:DualProblem}) can be solved iteratively. We first fix $\lambda$ and use any line search method to solve (\ref{Eqn:DualProblem}), and we obtain $N_{\text{T}}^{\ast}$. Next, we update the dual variable $\lambda$ as
\begin{align}\label{Eqn:DualUpdates}
\lambda_{l+1} = \left[\lambda_l - \alpha_l \left( R \left(N_{\text{T}}^{\ast},N_{\text{cpb}}\right) - R_0 N_{\text{S}} \right)\right]^{+},
\end{align}
where the operator $[x]^{+}$ denotes $\max(x,0)$ and $\alpha_l$ is the step size of the $l$th iteration. We keep iterating until the stopping condition is satisfied. The algorithm terminates when the relative change in $N_{\text{T}}$ between two iterations is less than some tolerance, i.e, $|N_{\text{T}}(l+1) - N_{\text{T}}(l)|/N_{\text{T}}(l) \leq \Delta$ \cite{NonlinearBazaraa}. 

The proposed algorithm, CLOEE, is summarized under the heading Algorithm~\ref{Algorithm}. We identify three scenarios. First, if the throughput at $N_{\text{T}}^{\text{EE}}$ satisfies the rate constraint, i.e., $R(N_{\text{T}}^{\text{EE}},N_{\text{T}}^{\text{cpb}}) > R_{0} N_{\text{S}}$, then $N_{\text{T}}^{\text{EE}}$ is obtained in a single-step using (\ref{Eqn:EEwoConstraints}). This occurs often for short to medium link distances. Second, if the throughput at $N_{\text{T}}^{\text{EE}}$ does not satisfy the rate constraint, but $N_{\text{T}}^{\text{THR}}$ satisfies it, i.e., $R(N_{\text{T}}^{\text{EE}}) < R_0 N_{\text{S}}$ and $R(N_{\text{T}}^{\text{THR}}) > R_0 N_{\text{S}}$, then we solve (\ref{Eqn:DualProblem}). In the link adaptation, this typically occurs during the mode transitions. Lastly, for long link distances, when there is no $N_{\text{T}}$ that satisfies the rate constraint at $N_{\text{cpb}}$, we assign $N_{\text{T}}^{\text{THR}}$ to $N_{\text{T}}$.




Figs.~\ref{Figure:EEThrVSPayload}(a)-(b) illustrate the dependence of the energy efficiency and throughput on the PSDU frame size. The points $N_{\text{T}}^{\text{EE}}$, $N_{\text{T}}^{\text{THR}}$, and $N_{\text{T}}^{\ast}$ are also shown. The results are given for a link distance of $8.4$~meters, where error probability is high. Rates with $N_{\text{cpb}} \in \{1,2,4\}$ perform very poorly, whereas the higher orders achieve better performance. Note that at a shorter distance, the order of these curves can be different and the performance depends on the link distance, see Figs.~\ref{Figure:DistanceVsKPI}(c)-(d).

\section{Numerical Results}\label{Section:Simulations}
In this section, we evaluate the performance of CLOEE. For comparison, we also simulate numerous static strategies and an exhaustive search approach. The performance results presented in this section are obtained by using Matlab. The channel characteristics of WBANs vary with the patient’s attributes (e.g., gender, weight, fat), radiation pattern that shapes the specific absorption rate (SAR), and patient motions. In this paper, the channel model in \cite{BAN09} is used to characterize the propagation environment in the 3.1-10.6~GHz band, for which the path loss is  \cite{BAN09} $L(d)  = a \cdot \log_{10}(d)  + b + \chi$, where $a$ and $b$ are constants, $d$ is the distance in millimeters, and $\chi$ is a Gaussian distributed random variable with a zero mean and a variance $\sigma^2$. Typical values for a hospital room are $a = 19.2$, $b = 3.38$, and $\sigma = 4.40$  \cite{BAN09}. The noise spectral density is taken as $-174$~dBm/Hz, $W_{\text{rx}}$ as $ 499.2$~MHz, noise figure as $10$~dB, and implementation margin as $5$~dB \cite{IEEE802156}. A non-coherent receiver and hard decision combining is considered, $\rho_r = 0$, and $\rho_c = 0$. For the energy consumption model, we have $\varepsilon_p = 20$~pJ, $P_{\text{COR}} = 10.08$~mW, $P_{\text{ADC}} = 2.2$~mW, $P_{\text{LNA}} = 9.4$~mW, $P_{\text{VGA}} = 22$~mW, $P_{\text{SYN}} = 30.6$~mW, $P_{\text{GEN}}= 2.8$~mW, and $T_{\text{ST}} = 400$~$\mu$s \cite{Seyedi10}.

Link distance and channel characteristics strongly influence the optimal PHY and MAC layer parameters for link adaptation. Figs.~\ref{Figure:DistanceVsKPI}(a)-(d) depict the link adaptation optimization as a function of the PSDU frame size and the number of pulses per burst when the link distance varies from one to ten meters. Figs.~\ref{Figure:DistanceVsKPI}(a)-(b) present the corresponding values of $N_{\text{T}}$ and $N_{\text{cpb}}$ obtained using CLOEE. It can be observed that for distances up to $7.5$~meters, longer frame sizes and higher data rates (few pulses per burst) provide the highest energy efficiency. For longer link distances, due to the lower SNR and thereby higher error probability, shorter frames sizes and more pulses per burst are preferred. For example, PSDU frame size decreases steeply after $7.5$~meters and the optimal number of pulses per burst soars to its maximum number of~$32$. In Figs.~\ref{Figure:DistanceVsKPI}(c)-(d), the performance of five static strategies, exhaustive search, and CLOEE are evaluated. The performance of CLOEE and exhaustive search overlap which demonstrates the advantage of CLOEE since its computational complexity is much smaller. As expected, CLOEE outperforms the static strategies in various performance metrics. For instance, the performance of $(N_{\text{cpb}},N_{\text{T}}) = (1,2616)$ and CLOEE are similar for short distances up to $4$~meters. Beyond this point, the performance of $(N_{\text{cpb}},N_{\text{T}}) = (2,2616)$ decays quickly, whereas CLOEE satisfies the QoS constraints up to $8.8$~meters, indicating its extended transmission range about a factor of two. Also, $(N_{\text{cpb}},N_{\text{T}}) = (32,2616)$ provides a very robust transmission, but it is highly inefficient for short distances since it achieves only $6.2$~Mbits/Joule, whereas CLOEE offers $57.5$~Mbits/Joule at the same distance, indicating close to an order of magnitude gain. 






\section{Conclusion}\label{Section:Conclusion}
Low energy consumption, reliability, and high energy efficiency are essential for the operation of WBAN devices. The energy efficiency of WBANs closely depends on the choice of PHY and MAC layer related parameters. We have proposed a cross-layer optimization for network energy efficiency maximization for IEEE 802.15.6 IR-UWB WBANs. In particular, we derived a closed-form expression to determine the optimal frame size subject to the minimum rate constraints; further a search is performed for optimal number of pulses per burst. Our simulation results demonstrate that the proposed CLOEE algorithm achieves the same performance as an exhaustive search and provides significant improvements in terms of energy efficiency (by an order of magnitude) and transmission range (by a factor of two) compared to the static strategies. In the future work, the proposed algorithm will be extended to approximate the frame error rate as a function of $P_b$ and $N_{\text{T}}$ to obtain explicit expressions for  $N_{\text{cpb}}$. As a final remark, the proposed CLOEE is universal and it can be applied to any network by updating the related frame parameters (e.g., size, sampling rate, and coding rate of the PHR, SHR, and PSDU), and bit error probabilities over the channel.


\appendix[Proof of Theorem~\ref{Eqn:TheoremEEQuasi}]\label{Appendix}
The second-order condition for a strictly quasiconcave function is that the second derivative needs to be non-positive at any point with zero slope \cite[p.~101]{OptBookBoyd}. Then, to prove the quasiconcavity of $\eta$, we need to show that 
\begin{align}\label{Eqn:QuasiconcavitySecondOrder} \nabla_{N_{\text{T}}} \eta = 0 \Rightarrow  \nabla_{N_{\text{T}}^2} \eta <  0. \end{align}
Let us define $A = P_{\text{SHR}} P_{\text{PHR}} (P_{\text{CW}})^{\frac{N_{\text{T}}}{n}}$, $\varepsilon_1 = \varepsilon_{\text{OH}} + \varepsilon_{\text{ST}}$, and $\gamma = \varepsilon_{\text{B}} N_{\text{T}} + \varepsilon_1$. Then, the gradient of $\eta$ is given by
\begin{align}\label{Eqn:GradientofEE} \frac{\partial \eta}{\partial N_{\text{T}}} = \frac{A (N_{\text{T}}^2 \frac{ \varepsilon_{\text{B}} \log(P_{\text{CW}}) }{n} + N_{\text{T}} \frac{\varepsilon_1 \log(P_{\text{CW}})}{n} + \varepsilon_1 ) }{\gamma^2}. \end{align}
Let us introduce $\beta = N_{\text{T}}^2 \frac{ \varepsilon_{\text{B}} \log(P_{\text{CW}}) }{n} + N_{\text{T}} \frac{\varepsilon_1 \log(P_{\text{CW}})}{n} + \varepsilon_1$. Note that $N_{\text{T}}^{\text{EE}}$ is the root of $\beta$, where the gradient becomes zero. The second derivative of $\eta$ can be expressed as 
\begin{align}\label{Eqn:QconcavityD2}
\frac{\partial^2 \eta}{\partial N_{\text{T}}^2} = \frac{A  \frac{\log(P_{\text{CW}})}{n}(\beta + 2 N_{\text{T}} \varepsilon_{\text{B}} + \varepsilon_1)}{\gamma^2} - \frac{A \beta (2  \varepsilon_{\text{B}})}{\gamma^3}.
\end{align} 
To prove that  (\ref{Eqn:QuasiconcavitySecondOrder}) holds true, we check the sign of $\frac{\partial^2 \eta}{\partial N_{\text{T}}^2}$ when the gradient is zero. Since $\log(P_{\text{CW}}) \leq 0$ for $0 \leq P_{\text{CW}} \leq 1$ and $\beta = 0$ at $N_{\text{T}}^{\text{EE}}$, we have $\frac{\partial^2 \eta}{\partial N_{\text{T}}^2} \leq 0$ which satisfies (\ref{Eqn:QuasiconcavitySecondOrder}). Thus, we prove that $\eta$ is strictly quasiconcave in $N_{\text{T}}$.

\section*{Acknowledgment}
Yang Liu's participation in this publication was made possible by NPRP grant \#6-415-3-111 from the Qatar National Research Fund (a member of Qatar Foundation). The statements made herein are solely the responsibility of the authors.

\bibliographystyle{IEEEtran}
\bibliography{IEEEabrv,Globecom16_july_finals}

\end{document}